\documentclass[floatfix, twocolumn]{article}

\usepackage{graphicx}
\usepackage{amsmath,amssymb,amsfonts}
\usepackage{textcomp}
\usepackage{verbatim}
\usepackage[normalem]{ulem}
\usepackage[numbers]{natbib}

\title{Small amplitude Dynamic AFM: quantifying interactions 
with different tip detection and excitation schemes in presence of additional resonances.}

\author{Luca Costa$^{1}$, Mario S Rodrigues$^2$ \\
\small{1- European Synchrotron Radiation Facility, 6 rue Jules Horowitz BP 220, 38043 Grenoble Cedex,France} \\
\small{2- CFMC/Dep. de F\'{\i}sica, Universidade de Lisboa,Campo Grande 1749-016 Lisboa, Portugal} \\
}

\begin{document}

\maketitle

\begin{abstract}
Quantifying the tip-sample interaction at the nanoscale in Amplitude Modulation mode AFM is challenging, 
especially when measuring in liquids.
Here, we derive formulas for the tip-sample conservative and dissipative interactions 
and investigate the effect that spurious resonances have on the measured interaction.
Both direct and acoustic excitation are considered. 
We also highlight the differences between measuring
directly the tip position or the cantilever deflection.
We show that, when probing the tip-sample forces, 
the acoustically excited cantilever behavior is insensitive to spurious resonances 
as long as the measured signal corresponds to the tip position, or if the excitation force is correctly taken into account.
Since the effective excitation force may depend on the presence of such spurious resonances, 
we consider the cases where the frequency is kept constant during 
the measurement so that the proportionality between excitation signal and actual excitation force is kept constant. 
With the present work we show the advantages that result 
from the use of a calibration method based on the acquisition of approach-retract curves. 
Optical beam deflection based AFMs benefit from the use of this calibration method, 
especially in presence of spurious resonances in the cantilever transfer function.
\end{abstract}

\let\thefootnote\relax\footnote{$^*$ Corresponding author: \textit{mmrodrigues@fc.ul.pt}}

\maketitle

\section{Introduction}
Dynamic AFM was introduced in the late 80s \cite{martin87a} as the natural evolution 
of the first Atomic Force Microscopes \cite{binning86}.
Thanks to its flexibility, Amplitude Modulation AFM (AM-AFM) \cite{garcia10} 
has become a successful dynamic operational scheme widely employed to characterize surfaces at the nanoscale. 
It has continuously evolved in terms of achievable lateral resolution and scan speed, 
producing impressive results both at the solid/gas interfaces in ambient conditions \cite{santos13} 
and at the solid/liquid interfaces \cite{tando2012}. 
A complete overview is given by references \cite{garcia10} and \cite{baro12}.
In AM-AFM, micro-sized cantilevers are conventionally excited 
at a frequency close to their first eigenmode. 
The oscillation amplitude of the tip is the feedback signal kept constant 
to obtain the sample morphology during the scan. 
Recently, the technique evolved in large amount of operational schemes based on the 
simultaneous detection of several cantilever eignemodes or harmonics of the tip displacement \cite{garcia_12,natraman2011}.

Evidently, the cantilever excitation plays a central role in AM-AFM.
Conventionally, a mechanical vibration in the cantilever holder 
is provided through the excitation of a small piezoelectric element (dither).
The setup is widely employed in many commercial and custom-made AFMs 
and permits measurements both in air and in liquids.
Despite the success of this method, 
the cantilever transfer function presents a forest 
of spurious peaks particularly when measuring in liquids. 
As a consequence, a quantitative estimation of the conservative and 
dissipative interaction between the probe and the sample is complicated. 
Moreover, it has already been observed that the motion 
of the cantilever base due to the acoustic excitation is not negligible in situations where the Q-factor is low,
a typical situation when measuring in liquids.
The same holds if the cantilever is not operated at a frequency close to its resonance frequency.
Several solutions have been proposed to overcome the presence of spurious 
peaks in the cantilever transfer function. 
One possibility is the development of custom-made liquid cells 
which limit the presence of spurious excitations \cite{2maali06} . 
Another possibility is the direct excitation of the tip bypassing 
the conventional piezoelectric excitation.  
In particular, in the last decade scientists have introduced magnetic \cite{han96,revenko00}, 
capacitive \cite{umeda10} and photothermal \cite{ratcliff98,ramos06} actuation schemes.

The differences between the direct excitation of the tip and the 
conventional dither excitation 
have already been studied and reported \cite{garcia07,raman07}. 
Consistent efforts have been done to properly quantify 
the conservative and dissipative interactions when using acoustically 
excited cantilevers in liquids \cite{raman11,mugele10}. 
The proposed methods have been successfully employed for 
AFMs based on an optical beam deflections scheme which provides a measurement of the tip bending angle\cite{meyer88b}.

Here we report a general study of the dynamics of acoustically 
excited cantilevers and lay-down formulas for deriving the conservative 
and dissipative interactions considering different types of
detection methods.

This results in a direct estimation of 
the conservative and dissipative interactions between the 
AFM probe and the specimen when employing Fabry-Perot 
interferometers as detection method \cite{rugar89,hoogenboom05}.
A second issue that is addressed in the present work is the consistent advantages 
that result from the use of approach-retracts curves as calibration method, 
as done with the Force Feedback Microscope\cite{io12,io13,mario13}, 
versus the standard characterization of the cantilever transfer function. 
With the described procedure, 
it would be possible to ignore the effects of spurious resonances when using optical beam deflection based AFMs.

Firstly, we derive simple equations to quantify the conservative and dissipative interactions 
in the small oscillation amplitudes regime assuming to know the tip position, 
which is the case of AFMs employing fiber-optic based interferometer.
Then, we focus on the problem of the cantilever motion coupled with an 
additional oscillator which may be represented by a resonance of the liquid cell.
Both cases where the cantilever is coupled to an extra oscillator 
at its base or directly at the tip position are considered. 
Finally, we report a description of the cantilever dynamics 
when the tip bending angle is measured, as in
conventional optical beam deflection operational scheme. 
The calibration procedure based on the acquisition 
of approach-retract curves and the derived formulas 
have been used to characterize a tip-sample electrostatic interaction, 
even in presence of spurious resonances in the cantilever transfer function.

\section{General formula for small oscillation amplitudes}

In this section we will produce two general formulas for the interaction stiffness $k_i$ and damping $\gamma_i$
without assuming the whole system has a specific transfer function dominated by the cantilever transfer function.
It will be assumed only that all the forces that are present are additive.

Consider a point mass which is being acted by two forces. One force given by $F_y(t)$ and another by $F_x(t)$.
The two forces add together to make a resultant force $F_r(t)$, which determines the motion of the mass:
\begin{equation} \label{eq:x}
x(t)= A \cos(\omega t+\phi)
\end{equation}
Which from Newton's second law means:
\begin{equation}
F_r=-m A \omega^2 \cos(\omega t+\phi) \label{eqfr}
\end{equation}
If we are interested to know $F_y(t)$, in the case where the forces are additive, 
we can put $F_y(t)=F_r(t)-F_x(t)$ i.e.:
\begin{eqnarray}
 F_y(t)&=&A_r \cos(\omega t + \phi)-A_x \cos(\omega t)
\end{eqnarray}
Where $A$ is the amplitude of the force identified with its respective subscript.
\noindent $F_x$ can be rewritten as:
\begin{equation}
 F_x(t)=A_x [\cos(\omega t +\phi) \cos(\phi)+\sin(\omega t +\phi) \sin(\phi)]
\end{equation}
From which we conclude:
\begin{equation}
 F_y(t)=[A_r-A_x \cos(\phi)]\cos(\omega t+\phi)-A_x \sin(\phi) \sin(\omega t +\phi)
\label{eq:blast_luca}
 \end{equation}

Consider that the force $F_y(t)$ has two contributions: 
one is a restoring force, i.e. an elastic contribution $F_{el}$ and the other is a damping force $F_{damp}$.
The elastic force is directly proportional to the position of the moving mass, 
whereas the damping is directly proportional to the first derivative of the position i.e. to its velocity.
Let us define $k$ as being the proportionality constant between the force and the position
and $\gamma$ the proportionality constant between the damping force and the speed of the mass.
Hence:
\begin{equation}
 F_y(t)=F_{el}(t)+F_{damp}(t)=-k x(t)-\gamma \frac{dx(t)}{dt}
 \label{eq:last}
\end{equation}
\noindent which means
\begin{equation}
 F_y(t)=-k A \cos(\omega t+\phi)+\gamma \omega A \sin(\omega t+\phi)
 \label{eq:last_luca}
\end{equation}
\noindent comparing equation \ref{eq:last_luca} 
with equation \ref{eq:blast_luca} gives:
\begin{eqnarray}
-k A&=&A_r -A_x \cos(\phi) \nonumber \\ 
\gamma \omega A &=& -A_x \sin(\phi)
\end{eqnarray}
\noindent If the position $x(t)$ of the mass is known, then the total force $F_r(t)$ to which 
the moving mass is submitted to, (\ref{eqfr}) is also known:
\begin{eqnarray}
k&=&m \omega^2 +\frac{A_x}{A} \cos(\phi) \nonumber \\
\gamma &=&-\frac{A_x}{\omega A} \sin(\phi) 
\end{eqnarray}
Now suppose that the proportionality constant $k$ between the force and the position depends on an external 
parameter $z$, suppose that it can be set $k=k_0+k_i(z)$
and that in the same way it can be set $\gamma=\gamma_0+\gamma_i(z)$.
Then in such case:
\begin{eqnarray}
k_i(z) &=&-k_0 +m \omega^2 +\frac{A_x}{A} \cos(\phi) \nonumber \\
\gamma_i(z) &=&-\gamma_0 -\frac{A_x}{\omega A} \sin(\phi) \label{eq:kigi1}
\end{eqnarray}
Assume $m$, $k_0$ and $\gamma_0$ are all unknown constants. 
These constants may have effective values that depend on the frequency, hence we will consider solely
the case where the frequency of motion $\omega$ is constant at all times, 
which is the common situation in AM-AFM.
Suppose also to know $k_i(\infty)=0$ and $\gamma_i (\infty) = 0$. 
Then, by setting the above equations equal to zero we obtain:
\begin{eqnarray}
 \frac{A_x^\infty}{A^\infty}& =& \sqrt{(k_0-m \omega^2)^2+\gamma_0^2 \omega^2} \equiv a\nonumber \\
 \phi^\infty& =&ArcTan\left( \frac{\gamma_0 \omega}{k_0-m \omega^2} \right) \label{eq:adelta}
\end{eqnarray}
\noindent The superscript infinity means that those constants are evaluated far from the surface.
In practice, due to the squeeze film effect, they should be evaluated before 
the short range forces but at few nanometers from that. 
We can then write a final relationship:
\begin{eqnarray}
k_i(z) &=& a[n \cos(\phi)-\cos(\phi^\infty)] \nonumber \\
\gamma_i(z)& =&\frac{a}{\omega} [\sin(\phi^\infty) -n \sin(\phi) ] \label{eq:final1}
\end{eqnarray}
\noindent where we have put:
\begin{equation}
 n\equiv \frac{A_x A^\infty}{A A_x^\infty}
\end{equation}
The amplitudes of excitation and tip motion have been normalized to $n$ which is then one far from the surface.
As long as the excitation force $A_x$ remains directly proportional to the excitation signal, 
it is not necessary to know its actual value,
nor how is the mass actually excited. 
It does not matter if the mass 
is excited by displacing the base of the cantilever or trough a direct actuation on the tip
or a combination of both, as long as they remain proportional to the excitation signal.
This formula is valid regardless of the presence of spurious peaks and or squeeze film effects, 
provided it can be assured the position is given in very good approximation by the expression in \ref{eq:x}.

A question that could arise is whether there is any difference in having spurious peaks or not?
$F_x(t)$ is the force acting on the moving mass.
This force depends on the excitation signal that can be controlled. 
Spurious peaks will affect the ratio between $F_x$ and the excitation signal as well as the phase between them.
We will come back to this later on, and consider specifically one spurious peak.
If the cantilever spectrum has well defined transfer function, then it is straightforward
to  obtain all constants above from that transfer function.
If the spectrum however is deformed by spurious resonances,
then that means that constants $a$ and $\phi^\infty$ cannot be evaluated from a simple analysis of the spectrum.
This does not mean however, that equations \eqref{eq:final1} are incorrect.
An important note is that even if the resonance curve is calibrated close to the sample, 
we assume that $k$ is the spring constant of the cantilever, so that a resonance frequency different from
the natural frequency is accounted by only trough a rescaling of the effective mass and quality factor.
Whereas, if $a$ and $\phi^\infty$ are calibrated, the cantilever spring constant is not 
fixed to any value. 

The above assumption requires in practice that the interaction is linear. 
This can be assured by keeping the amplitude of oscillation small enough.
It is assumed the measurement is the position of the particle to which the forces are applied.
Most of the AFMs employ however optical beam deflection schemes \cite{meyer88b}, 
providing the measurement of the tip bending angle and not the tip position.
If the cantilever is directly excited so that the base of the cantilever is not displaced
\cite{han96,revenko00,umeda10,ratcliff98,ramos06}, then equations \eqref{eq:final1} are still valid, 
because in that case the deflection is indeed proportional to the position, 
implying that the measured signal is directly proportional to the total force applied on the tip.
If the tip position is not measured and the cantilever is not directly excited, 
then equations \eqref{eq:final1} do not hold, particularly off the resonance or at the resonance but with small $Q$ factor.

\section{Coupling with one extra resonance}
Consider now that the cantilever is coupled to another oscillator giving rise to one extra peak in the spectrum.
The cantilever may be coupled in different ways.
Let us consider firstly the case when the coupling is at the cantilever base, described by the equations:
\begin{eqnarray}
\label{equ:new_luca}
m_l \ddot{x_l}+k_l(x_l-x_b)+\gamma_l \dot{x_l} &=&0 \\
m_s \ddot{x_s}+k_l(x_s-x_l)+k_s(x_s-x_0)+\gamma_s \dot{x_s} &=0&
\end{eqnarray}
This case corresponds to situations where additional resonances 
in the cantilever transfer function are due to the dither piezoelement 
in the cantilever holder or to the cantilever holder itself.
Here the subscript $l$ stands for lever and $s$ for spurious.
This system becomes straightforward to solve, considering spurious peaks 
at frequencies close to the cantilever frequency
and assuming that nothing else in the system is as soft as the cantilever.
If the frequency is comparable to the cantilever resonance frequency,
then the ratio $k_s/m_s$ is also comparable.
If both the spurious motion is comparable to the cantilever motion, 
then the situation is such that the term $k_l(x_s-x_l)$ is negligible when compared to the other terms.
This implies that the spurious motion is insensitive to motions of the cantilever
i.e. the spurious motion is the same regardless of the cantilever vibrations.
Thus, this motion depends only on the excitation signal and is the same regardless of the cantilever being present or not.
As a matter of fact, if the set up cannot compel with this requirement,
then the static spring constant of cantilever could not be used for 
any quantitative evaluation of the interaction, 
because in this case the deflection due to the tip-sample forces would not 
be directly proportional to the cantilever spring constant.
As a consequence, the above condition/simplification corresponds to the real situation.
By accepting this, equation \eqref{equ:new_luca} can be rewritten as:
\begin{eqnarray}
m_l \ddot{x_l}+k_l(x_l-x_s)+\gamma_l \dot{x_l}  &=&0 \nonumber \\
m_s \ddot{x_s}+k_s(x_s-x_0)+\gamma_s \dot{x_s} &=& 0
\end{eqnarray}
The second equation has a very well know steady state solution.
A spurious oscillation will occur with amplitude: $A_s = x_0/\sqrt{(k_s-m_s \omega^2)^2+\gamma_s^2 \omega^2}$
and phase $\phi_s=\textrm{ArcTan}[\gamma_s \omega/(k_s-m_s \omega^2)]$.
The problem then simplifies and summarizes in the solution of:
\begin{eqnarray}
m_l \ddot{x_l}+k_l(x_l-A_s \cos(\omega t+\phi_s))+\gamma_l \dot{x_l} &=&0 \label{equ:def1}
\end{eqnarray}
Which has the steady state solution given by:
\begin{eqnarray}
A_l = \frac{k_l A_s}{\sqrt{(k_l-m\omega_ 2)^2+\gamma_l^2 \omega^2}}
\end{eqnarray}
and phase respective to the excitation signal given by:
\begin{eqnarray}
\phi_l = -\textrm{ArcTan}\left(\frac{\gamma_l \omega}{k_l-m_l \omega^2} \right)+\phi_s
\end{eqnarray}
Clearly, at a given frequency, the effect of a spurious resonance of the kind considered here is that the excitation signal is amplified by the 
spurious resonance ($A_s$) and a phase lag is introduced. 
However, if the frequency is kept constant, this amplification factor 
and phase lag remain constant.
The quantity that matters $a=A_s/A_l$ remains unchanged.  
The effect of this spurious peak is that, depending on its resonance, 
the ratio between the actual excitation and the supplied excitation will change.
We then conclude that this type of spurious peak do not affect the measured tip-sample interaction.

Consider now that the cantilever is coupled not at the base but at the tip position:
\begin{eqnarray}
m_l \ddot{x_l}+k_l(x_l-x_b)+\gamma_l \dot{x_l} +k_c (x_l-x_c)  &=&0 \nonumber \\
m_c \ddot{x_c}+k_c(x_c-x_l)+\gamma_c \dot{x_c}+... &=0&
\end{eqnarray}
This case corresponds to a situation where additional resonances 
in the cantilever transfer function are due to the fluid borne excitation as described in reference \cite{raman11}.
Here, $k_c$ represents a restoring force that may be induced by the fluid.
Contrary to the previous case, now it is preferable to have a small value for constant $k_c$.
We will not be able to solve the system above but we can rewrite the first equation as:
\begin{eqnarray}
m_l \ddot{x_l}+(k_l+k_c) x_l +\gamma_l \dot{x_l}  &=& x_b k_l+k_c x_c  
\end{eqnarray}
Similarly to the previous case, both $x_b$ and $x_c$ are directly proportional to the excitation signal, 
so they add to give $A_x$ also proportional to the excitation signal.
We may not be able to tell how much is $A_x$. 
However, it is not difficult to see that in this case equation \ref{eq:final1} still holds, 
except that now $k_0=k_l+k_c$. 
Due to the coupling, the cantilever has a different effective and unknown spring constant.
If constants $a$ and $\phi^\infty$ are determined directly by using a known interaction, 
then the change in cantilever stiffness is directly taken into account.
The constants $a$ and $\phi^\infty$ can be calibrated by measuring tip position, 
oscillation amplitude and phase using a tip-sample known force or using for example a tip-sample approach curve
and fitting $F=-\int k_i dz$ with equation \ref{eq:final1} as in references [22]
and [23]. 
The tip position provides the static force $F$, whereas amplitude and phase provide $k_i$ trough the calibrated constants $a$ and $\phi^\infty$.
An electrostatic force may be easily employed in air for this purpose. 
In liquids it can be done for example by measuring tip-sample approach curves, assuming the static force to be conservative.

An intrinsic limit to any calibration method comes 
from the time dependence of the calibration parameters.
If the liquid cell has an open architecture, 
the evaporation of the liquid drop affects directly these parameters 
which are needed to quantify the interaction. 
They should then be evaluated continuously.

\begin{figure}[ht]
\begin{center}
\includegraphics[width=6cm]{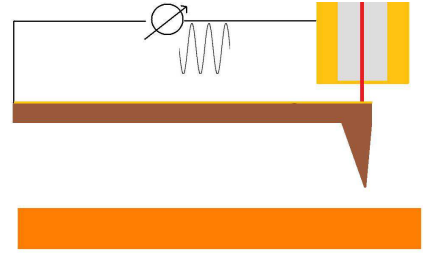}  	
\end{center}		
\caption{
Setup employed for the direct excitation of the tip. 
A gold coated optic fiber is used to measure the tip position 
and to apply an electrostatic excitation of the conductive cantilever at a frequency close to resonance.}
\label{fig:figure1} 
\end{figure}

We employed the calibration method requiring the acquisition of tip-sample 
approach curves to compare measurements with a conventional 
acoustic excitation  and a direct capacitive excitation of the tip.
We coated an optic fiber, permitting the measurement of the tip position, 
with 30 nm thickness of gold at its end and 300 nm thickness of gold on its borders as shown in figure \ref{fig:figure1}.

\begin{figure}[!ht]
\begin{center}
\includegraphics[width=9cm]{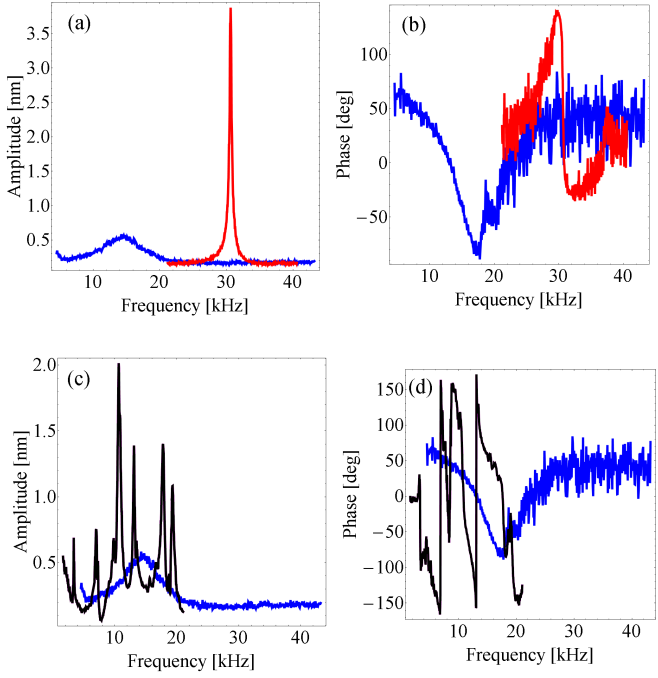}  	
\end{center}		
\caption{
Excitation of the tip in air and in liquid with different actuation methods:
a) Amplitude and b)phase of the tip directly excited in air (red) and (liquid). 
c) Amplitude and d) phase of the tip excited in liquid with a capacitive actuation (blue) 
and with a conventional piezoelectric excitation.}
\label{fig:figure2} 
\end{figure}

We applied a harmonic oscillating signal $V(\omega)$ around a given $\Delta V_0$. 
The force applied on the tip is given by:
\begin{equation}
F = -\frac{1}{2} \frac{\partial C}{\partial z} \left[\Delta V_0 + V(\omega)\right]^2 
\end{equation}
where $C$ is the capacitance resulting from the plates constituted by the
cantilever backside and the bottom of the gold coated optical fiber.
Figures \ref{fig:figure2}a and b show the amplitude and the phase 
of the excited cantilever in air (red) and in deionized water (blue). 
Typical values for $\Delta V_0$ are 5 V in air and 500 mV in liquid, 
whereas the amplitude of the harmonic modulation is 100 mV. 
The tip-fiber distance is 10 $\mu$m. 
At this distance, the force acting on 40 $\mu$m wide cantilevers is in the order of 10 nN/V 
in static conditions in air. In water no capacitive force acting on the cantilever was observed in 
quasi-static conditions, whereas figure \ref{fig:figure2}a shows the presence of capacitive actuation at frequencies higher than 1 kHz. 
The diameter of the optic fiber is 125 $\mu$m. 
The cantilevers have nominal spring constant of 0.8 N/m. 

Figures \ref{fig:figure2}a and b show the amplitude and the phase of the cantilever 
in liquid with a direct excitation of the tip (blue) and with a conventional 
piezoelectric excitation (black), showing the presence of spurious resonances.

\begin{figure}[ht]
\begin{center}
\includegraphics[width=9cm]{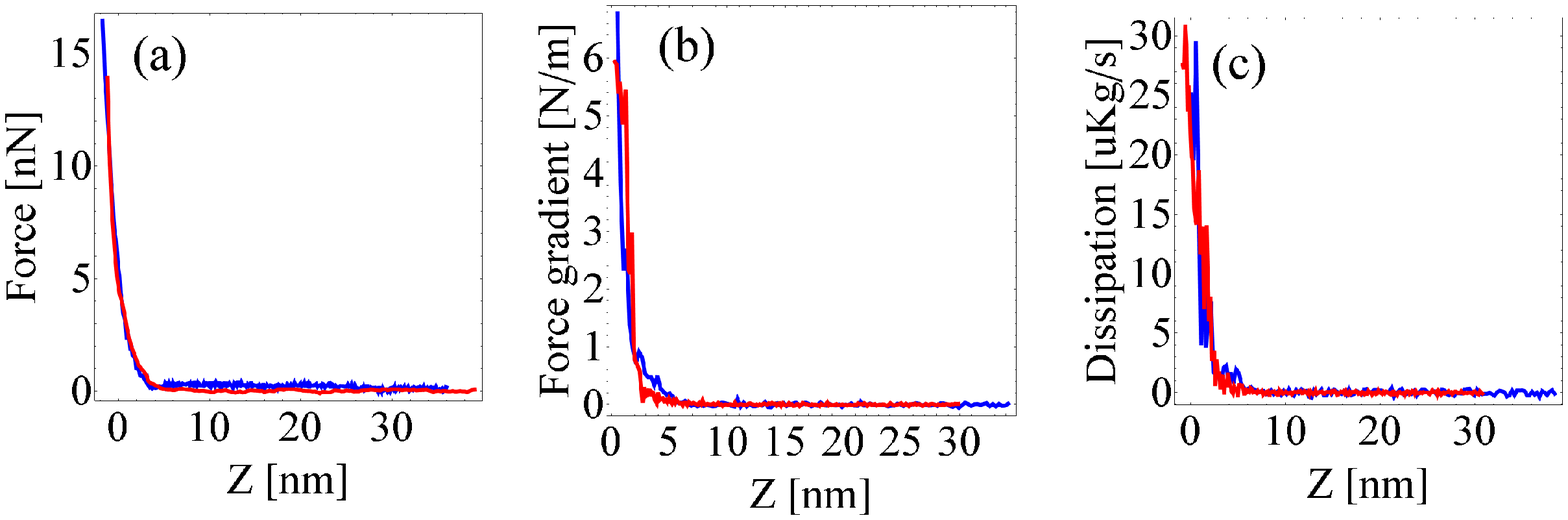}  	
\end{center}		
\caption{
Interaction at the mica/deionized water interface with electrostatic excitation of the tip (blue) and 
conventional piezoelectric excitation (red): 
a) Static force. b) Force gradient. c) Dissipation.}
\label{fig:figure3} 
\end{figure}

We acquired approach force curves in Force Feedback Mode \cite{io12,io13} at the mica/deionized water interface.
The oscillation amplitude imposed to the tip is 0.3 nm.
The amplitude and the phase were recorded and converted into conservative 
and dissipative interactions using equations \ref{eq:final1} and the equality $F=- \int k_i dz$, 
assuming the static force to be fully conservative. 
Constants $a$ and $\phi^\infty$ were determined.
Then, we performed the same measurements exciting 
the tip with the same oscillation amplitude and with the conventional piezoelectric excitation. 
Even in presence of spurious peaks, we converted
the amplitude and phase into conservative and dissipative interactions, 
providing new $a$ and $\phi^\infty$ parameters.
Figure \ref{fig:figure3}a shows the static force measured while the tip was electrically 
excited (blue) and piezoelectric excited (red). 
The elastic force gradient and the dissipation are shown in 
figure \ref{fig:figure3}b and \ref{fig:figure3}c. 
Providing the same results in terms force gradient and dissipation, 
the main difference between the two excitation methods is 
then solely given by the calibration parameters $a$ and $\phi^\infty$ 
evaluated through the equality $F=- \int k_i dz$.

In the case of electrostatic excitation, 
we obtained $\phi^\infty$ = 0.05 rad, $a$ = 2.2N/m.
For the conventional acoustic excitation, 
we have $\phi^\infty$ = 0.006 rad, $a$ = 0.06N/m.

\section{Measurement based on deflection angle}

If the measurement is not directly proportional to the position of the cantilever 
but only proportional to its deflection, 
then the displacement of the base has to be added to the measured signal.
We can start by considering equation \ref{eq:kigi1} and simply recomputing both $\cos(\phi)/A$
and $\sin(\phi)/A$.
The position of the moving mass is given by its deflection plus the motion of the base:
$A \cos(\omega t +\phi)=A_m \cos(\omega t + \phi_m) + A_b \cos(\omega t+\phi_b)$. 
\noindent Where $A_m$ is the amplitude of the measured signal and 
$A_b$ the oscillation amplitude of the cantilever anchoring point.
Equation \ref{eq:kigi1} becomes:
\begin{eqnarray}
 k_i=-k+m \omega^2 + \frac{A_x \left[A_m \cos(\phi_m)+A_b \cos(\phi_b)\right]}{A_m^2+A_b^2+2 A_b A_m \cos(\phi_m-\phi_b)} \nonumber \\
 \gamma_i=-\gamma_0-\frac{1}{\omega} \frac{A_x \left[A_m \sin(\phi_m)+A_b \sin(\phi_b)\right]}{A_m^2+A_b^2+2 A_b A_m \cos(\phi_m-\phi_b)} \nonumber \\
 \label{eq:comp}
\end{eqnarray}
Note that equations \ref{eq:comp} are exactly equal to equations \ref{eq:final1}
if one of the following is true:
\begin{enumerate} 
 \item The measured signal corresponds to the position of the moving tip, 
 in which case the terms with $A_b$ do not appear in the equation.
 \item It is assured that the base of the cantilever does not move (direct excitation), in which case $A_b=0$.
 \item It is a very good approximation to state that $A_m>>A_b$ ($A_b$ is negligible compared to $A_m$). 
 which is true close to the resonance frequency when the cantilever quality factor is large enough.
\end{enumerate}
Interestingly, it means that to be able to use the equations described in the first part (\ref{eq:final1}), 
it is convenient to either make a direct excitation of the cantilever 
or to measure directly a signal proportional to its position. 
In the last case, it becomes irrelevant how the moving mass is excited.
There is one particular case worth noting when the excitation force is solely 
due to the cantilever base displacement.
In this case it is $A_x=k A_b$ and $\phi_b=0$). 

Let us also define the proportionality constant $r$ such that $n r A_m=A_x$.
Here, $r$ is a constant such that in the absence of tip-sample interactions the normalized amplitude ($n$) is one.

Equations \ref{eq:comp} then become:
\begin{eqnarray}
 k_i&=&k\left(\frac{m \omega^2}{k} -\frac{k(k+n r \cos(\phi_m))}{k^2+n^2 r^2+2 k n r \cos(\phi_m)} \right)  \nonumber \\
 \gamma_i&=&\frac{k}{\omega}\left(-\frac{\omega \gamma_0}{k} -\frac{k n r \sin(\phi_m)}{k^2+n^2 r^2+2 k n r \cos(\phi_m)} \right) \nonumber \\
 \label{eq:comp_luca}
\end{eqnarray}
Far from the sample surface there are no short range sample interactions and equations 
\eqref{eq:comp_luca} must evaluate to zero.
We can solve equations \eqref{eq:comp_luca} to find $\phi_m^\infty$ and $r$ that are constants:
\begin{eqnarray}
 \frac{A_x^\infty}{A_m^\infty}& \equiv &r=k \sqrt\frac{\gamma^2 \omega^2+(k-m w^2)^2}{\gamma^2 \omega^2 + m^2 \omega^4} \nonumber \\ 
 \phi_m^\infty&=&\textrm{ArcCos}\left( \frac{\omega (\gamma^2+m(m \omega^2-k))}{\sqrt{(\gamma^2+m^2 \omega^2)(\gamma^2 \omega^2+(k-m \omega^2)^2)}} \right) \nonumber \\
 \label{eq:constants}
\end{eqnarray}
Then, the tip-sample interaction is given by:
{\small
\begin{align}
 k_i=k\left(\frac{k(k+r \cos(\phi_m^\infty))}{k^2+r^2+2 k r \cos(\phi_m^\infty)} -\frac{k(k+n r \cos(\phi_m))}{k^2+n^2 r^2+2 k n r \cos(\phi_m)} \right) \nonumber \\
 \gamma_i=\frac{k}{\omega}\left(\frac{k r \sin(\phi_m^\infty)}{k^2+r^2+2 k r \cos(\phi_m^\infty)} -\frac{k n r \sin(\phi_m)}{k^2+n^2 r^2+2 k n r \cos(\phi_m)} \right)  \nonumber\\
 \label{eq:comp_luca2}
\end{align}}
This highlights the fact that if the cantilever spring constant $k$ is known, 
then the evaluation of either $\gamma$ and $m$ or $r$ and $\phi^\infty$ is needed to quantify the interaction.
The first couple can be found trough a resonance curve if it is well defined in the spectrum.
In that case one can compute $r$ from the expression above.
The second couple can be found by taking a calibration curve using a known 
force or using for example an approach curve and fitting $F=- \int k_i dz$ as in 
[22,23]. 
In air it is straightforward to calibrate these parameters using an electrostatic force.
If it is not possible to obtain these values from the resonance curve because 
the spectrum is very deformed, that does not mean these equations do not hold.

Finally, some considerations regarding the case where $A_x \neq k A_b$. 
First note that this situation poses no difficulty in the case where the tip motion is measured instead of the cantilever deflection.
It is more complicated in the case of beam deflection because equation \ref{eq:comp} is more difficult to analyze. 
The values of $A_x$ and $A_b$, or their ratios, 
are not acquired from a conventional analysis of the transfer function nor from fitting the Brownian motion of the cantilever.
Also note that this is exactly the case where piezoelectric actuation is used and liquid borne excitation is not negligible.
However, one can use the equality $F=\int \nabla F dz$ calibrate either equation \ref{eq:comp} or \ref{eq:comp_luca2} at a given frequency.

\begin{figure}[htb]
 \includegraphics[width=\linewidth]{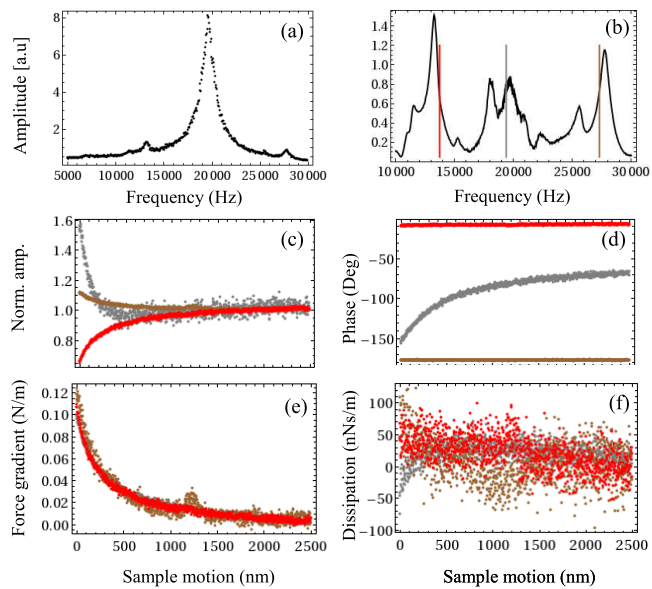}
 \caption{Characterization of a tip-sample electrostatic interaction at resonance 
 (gray) and off resonance (red and brown) a) Cantilever Brownian motion; b) spectrum of the excited cantilever; 
 c) normalized amplitude at three selected frequencies indicated in the spectrum; 
 d) measured phase (offset so that at infinity it gives $\phi_m^\infty$); e) and f)
 force gradient and dissipation measured at the three selected frequencies.
 }
\end{figure}

To illustrate the situation described above, 
we have prepared a cantilever holder that excites the cantilever by a combination of rotation and translation. 
In this case $A_x \neq k A_b$. 
Furthermore, this particular holder shows many spurious resonances.
Three approach curves were taken at selected arbitrary frequencies as shown in figure 4b. 
we applied an electrical potential difference between 
a conductive cantilever and a conductive sample which results in an electrostatic interaction.
We have used equations \ref{eq:comp_luca2} to account for the interaction. 
As shown in figure 4e, all the measurements provide the same force gradient 
regardless of the presence of spurious resonances shown in figure 4a, 
whereas the dissipation remains constant and approximately equal to zero.
The static force was used to obtain all the unknown parameters.
This result clearly shows that a calibration method is advantageous compared to a conventional analysis of the transfer function.

Interestingly, the values of $r$ and $\phi^\infty$ that 
result from the calibration are those that would be expected if the spurious peaks were not there. 
Notice in figure 4d the phase close to 0 Deg, -90 Deg and -180 Deg, when the cantilever
is respectively excited before, close to, and far from the resonance frequency.
To ease the task of calibrating the cantilever, $r$ and $\phi^\infty$ were initially guessed from equations \ref{eq:constants}, 
where the constants therein were obtained from the analysis of the tip Brownian motion shown in fig 4a.
The resulting $r*$ and $\phi^{\infty*}$ are shown in table 1 and compared with $r$ and $\phi^\infty$ 
obtained from the analysis of the Brownian motion.
This result means that the change in amplitude and phase due to the interaction force 
were about the same, regardless of the presence of spurious peaks.

\begin{table}
 \begin{tabular}{cccc}
  $f_x$ & 13734 Hz & 19393 Hz & 27315 Hz\\
  $r$ & 1.05 N/m& 0.059 N/m & 0.48 \\
  $r*$ &1.05 & 0.056 & 0.4 \\
  $\phi^\infty$ &-8 Deg &-66Deg & -178Deg \\
  $\phi^{\infty*}$ &-8 Deg &-66Deg &-177Deg \\
 \end{tabular}
\caption{Comparison between the calibration based on matching the integral of the force gradient
to the force, and using directly the constants determined from fitting the Brownian motion of the tip.
The $^*$ corresponds to values obtained from matching the integral of the force gradient to the force.}
\end{table}

\section{Conclusions}
We have introduced a methodology to directly derive the conservative and dissipative interactions 
between the AFM probe and the sample in dynamic AFM experiments 
when small oscillation amplitudes of the tip are used and for different tip excitation and detection schemes. 
We considered both direct detection of the tip position 
(for example with Fabry-Perot interferometers) and classic optical beam deflection scheme, 
showing that the first method allows a much easier extraction of the conservative 
and the dissipative interactions without any knowledge of the effects of spurious resonances or the cantilever base motion. 
Even optical beam deflection based AFMs allows to extract 
the conservative and dissipative part of the interaction, 
performing the calibration procedure of the cantilever dynamics through 
the acquisition of tip-sample approach curves instead 
of the classic characterization through the cantilever transfer function.

Measuring directly the dip position greatly simplifies the analysis of the interaction. 
We have demonstrated in fact that the knowledge of the cantilever base motion 
is not necessary in order to quantify the conservative and dissipative interaction 
in dynamic AFM experiments in liquid if an interferometer is used to measure the tip position.
The presence of spurious peaks related to the liquid cells resonances in the cantilever 
transfer function is known to introduce problems in the evaluation of the cantilever properties such as mass, 
spring constant and damping. We have proposed a calibration method of the cantilever response 
that is not based on the measurement of its transfer function, 
but it is based on the acquisition the tip position/deflection, 
amplitude and phase during approach curves and consequently the assumption of the equality 
$F=- \int k_i dz$. The method has been employed 
to compare the conservative and dissipative interaction at the mica/deionized water interface measured 
with a conventional piezoelectric dither excitation and direct electrostatic excitation. 
The different setups provide the same evaluation of the force gradient, 
although the calibration parameters are different. 
We have pointed out the limits of this method, 
which requires a continuous calibration of the cantilever parameters if 
the measurement is performed with a setup subject to evaporation of the liquid drop.
This limit is not intrinsic to the calibration method suggested in this
work but is indeed common to any calibration procedure
if the evaporation of the liquid drop is not avoided.

Finally, we have highlighted the changes that have to be introduced 
into the formulas for optical beam deflection based AFMs 
and we have shown how to quantify the conservative and dissipative part of a tip-sample electrostatic 
interaction in presence of spurious resonances in the cantilever transfer function.

\section*{Acknowledgments}
Luca Costa acknowledges COST Action TD 1002. 
Mario S. Rodrigues acknowledges financial support from Funda\c{c}\~{a}o para a Ci{\^e}ncia e Tecnologia SFRH/BPD/69201/2010. 
The authors thank Fabio Comin for discussions and Irina Snigireva for the preparation of the gold coated optic fibers.

\end{document}